\documentclass[preprint]{emulateapj}

\usepackage{color}

\usepackage{amsmath}
\usepackage{amssymb}

\newcommand{\ra}[3]{$#1^\mathrm{h} #2^\mathrm{m} #3^\mathrm{s}$}
\newcommand{\dec}[3]{$#1^\circ #2' #3''$}
\newcommand{\err}[3]{$#1^{+#2}_{-#3}$}
  
\slugcomment{To appear in Astrophysical Journal Letters}

\shorttitle{Spectroscopic Confirmation of Coma UDGs}
\shortauthors{Kadowaki et al.}

\begin{document}


\title{Spectroscopy of Ultra-diffuse Galaxies in the Coma Cluster}


\author{Jennifer Kadowaki, Dennis Zaritsky, and R. L. Donnerstein}
\affil{Steward Observatory and Department of Astronomy, University of Arizona, Tucson, AZ 85719}
\email{jkadowaki@email.arizona.edu}




\begin{abstract}
We present spectra of 5 ultra-diffuse galaxies (UDGs) in the vicinity of the Coma Cluster obtained with the Multi-Object Double Spectrograph on the Large Binocular Telescope. 
We confirm 4 of these as members of the cluster, quintupling the number of spectroscopically confirmed systems. Like the previously confirmed large (projected half light radius $>$ 4.6 kpc) UDG, DF44, the systems we targeted all have projected half light radii $> 2.9$ kpc. As such, we spectroscopically confirm a population of physically large UDGs in the Coma cluster.
The remaining UDG is located in the field, about $45$ Mpc behind the cluster. We observe Balmer and Ca II H \& K absorption lines in all of our UDG spectra. By comparing the stacked UDG spectrum against stellar population synthesis models, we conclude that, on average, these UDGs are composed of metal-poor stars ([Fe/H] $\lesssim -1.5$). We also discover the first UDG with [OII] and [OIII] emission lines within a clustered environment, demonstrating that not all cluster UDGs are devoid of gas and sources of ionizing radiation.
\end{abstract}



\section{Introduction}
Ultra-diffuse galaxies (UDGs) are a class of spatially-extended, low surface brightness galaxy. 
To understand their physical nature and study them in the context of their environment, we require 
accurate distance measurements.

Spectroscopic redshifts have been measured for 4 galaxies categorized as UDGs since the renewed interest in extreme low surface brightness galaxies in 2015. Of the population discovered in proximity to the Coma Cluster, DF44 ($\alpha=$ \ra{13}{00}{58.0}; $\delta=$ \dec{26}{58}{35}) is the only UDG spectroscopically confirmed as a Coma member \citep{vanDokkum2015b}. Likewise, a spectroscopic redshift measurement places VCC 1287 ($\alpha=$ \ra{12}{30}{23.65}; $\delta=$ \dec{13}{56}{46.3}) in the Virgo Cluster \citep{Beasley2016} and DGSAT I ($\alpha=$ \ra{01}{17}{35.59}; $\delta=$ \dec{33}{31}{42.37}) in a low density filament of the Pisces-Perseus Supercluster \citep{Martinez-Delgado2016}, identifies UGC 2162 as the nearest UDG \citep{trujillo2017}. While DF44 and VCC 1287 are projected near their associated galaxy clusters, other apparent close associations can be misleading, as demonstrated by \cite{Merritt2016}.
All coordinates listed in this paper are given in J2000.

In addition to providing the critical distance measurement, spectra provide information on the current stellar populations and star formation history of these galaxies. For example, 
\cite{vanDokkum2015b} noted that the spectrum of DF44 is similar to that of early-type galaxies, with Balmer and G-band absorption lines, concluding that DF44 is a quiescent galaxy with no significant on-going or recent star formation. Such findings inform the development of models for the origin and evolution of these systems. We aim to determine if there does exist a class of physically extended UDGs, like DF44, in significant numbers in the cluster environment and if these galaxies are exclusively quiescent, old galaxies. In \S\ref{data} we describe the observations and data analysis. In \S\ref{results} we present the redshift measurements and interpret the spectra. Throughout we adopt a standard $\Lambda$CDM cosmology with $\Omega_m = 0.286$, $\Omega_\Lambda = 0.714$, and $H_0 = 69.6 \, \mathrm{km \, s^{-1}}$.


\section{Data}
\label{data}

We selected UDGs to observe from the \cite{vanDokkum2015a} sample of 47 candidates projected near the Coma Cluster. In particular, we selected
luminous ($M_g \lesssim -14.2$ mag) and large UDGs (with projected half light radii, $r_\mathrm{e}$, larger than 2.9 kpc assuming the UDG is in Coma) to focus on those analogous to DF44. Only 18 UDGs from the parent sample satisfy both the size and luminosity criteria. Finally, we had to require that potential targets have a nearby guide star brighter than $M_R<15$ that is suitably located for the guide probe. Our final sample contained 10 candidate targets, of which we observed six. We list the observed UDGs and relevant parameters in Table \ref{table:udg_target}.

On 2016 March 9 and 10 we observed in mostly clear conditions using the Multi-Object Double Spectrograph \citep[MODS;][]{Pogge2010} on the Large Binocular Telescope (LBT) prior to the binocular mode being operational.
Because UDGs have relatively flat surface brightness profiles, 
we opted to use a wide slit to increase the amount of light collected, at the expense of spectral resolution. We used a custom 2.4\arcsec-wide slit, twice as wide as the largest previously available long slit. 
We observed each target for 60 to 90 minutes, combining multiple 30 minute dithered exposures,
simultaneously using the G400L grating (400 lines mm$^{-1}$ blazed at $4.4^\circ$ centered on 4000\AA\  with a resolution of 1850) in the blue channel (3200-5800 \AA) and the G670L grating (250 lines mm$^{-1}$ blazed at $4.3^\circ$ centered on 7600\AA\  with a resolution of 2300) in the red channel (5800-10000 \AA). 
We do not use the red channel data due to higher sky brightness. 

We reduced the spectra using the MODS data reduction pipeline \citep{Pogge2010}. Reduction includes the standard steps of overscan correction, 2D bias subtraction, dark current subtraction, flat fielding, wavelength calibration, sky subtraction, flux calibration, and extraction. 
Because we use the MODS long-slit, over 80\% of the slit is clear sky, allowing for high S/N sky subtraction using the two-dimensional fitting available in the standard pipeline. Our final spectra have a S/N in the continuum ranging from 3 to 8 per pixel. These are therefore fairly low S/N spectra.
We obtained these redshift-quality spectra for six targets, but  exclude DF17 from further discussion because ultimately we were unable to measure a redshift.
%


\begin{table*}
\begin{center}
\caption{Our Spectroscopic UDG sample\tablenotemark{a}\label{table:udg_target}}
\begin{tabular}{ccccccccr}
\tableline
\\
UDG & RA & Dec & $R$\,\tablenotemark{b} & $\mu(g,0)$ & $r_\mathrm{eff}$ &$M_g$&$t_{exp}$	&$cz$\\
&(J2000)&(J2000)&(arcmin)&(mag arcsec$^{-2}$)&(kpc)&(mag)&(min)&(km s$^{-1}$)\\
\\
\tableline 
\\
DF03\tablenotemark{c} &\ra{13}{02}{16.5} &\dec{28}{57}{17} &66.9 &\err{24.5}{0.5}{0.5}  &\err{4.2}{1.2}{1.0} &\err{-15.0}{0.3}{0.2} &90	&$10150 \pm	37$ \\
DF07 &\ra{12}{57}{01.7} &\dec{28}{23}{25} &44.3 &\err{24.4}{0.5}{0.5}  &\err{4.3}{1.4}{0.8} &\err{-16.0}{0.2}{0.2} &90 &$6587 \pm	33$ \\
DF08 &\ra{13}{01}{30.4} &\dec{28}{22}{28} &32.6 &\err{25.4}{0.5}{0.5}  &\err{4.4}{1.5}{0.9} &\err{-14.9}{0.3}{0.3} &60	&$7051 \pm	97$ \\
DF17 &\ra{13}{01}{58.3} &\dec{27}{50}{11} &29.9 &\err{25.1}{0.5}{0.5}  &\err{4.4}{1.5}{0.9} &\err{-15.2}{0.3}{0.2} &90	&...\\
DF30 &\ra{12}{53}{15.1} &\dec{27}{41}{15} &88.8 &\err{24.4}{0.5}{0.5}  &\err{3.2}{0.9}{0.6} &\err{-15.2}{0.2}{0.2} &60	&$7316 \pm	81$ \\
DF40 &\ra{12}{58}{01.1} &\dec{27}{11}{26} &53.1 &\err{24.6}{0.6}{0.6}  &\err{2.9}{0.7}{0.5} &\err{-14.6}{0.2}{0.2} &60	&$7792	\pm 46$ \\
\\
\tableline
\end{tabular}
\tablenotetext{1}{The coordinates, surface brightness, effective radii, and absolute magnitude are quoted from \cite{vanDokkum2015a}.}
\tablenotetext{2}{The projected angular separation was computed from the center of the Coma Cluster ($\alpha=$ \ra{12}{59}{48.7}; $\delta=$ \dec{27}{58}{50}).}
\tablenotetext{3}{Because DF03 is located behind the Coma Cluster, the values for the effective radius and absolute magnitude have been recomputed using the new distance.}
\end{center}
\end{table*}





\begin{figure}
\epsscale{1.0}
\plotone{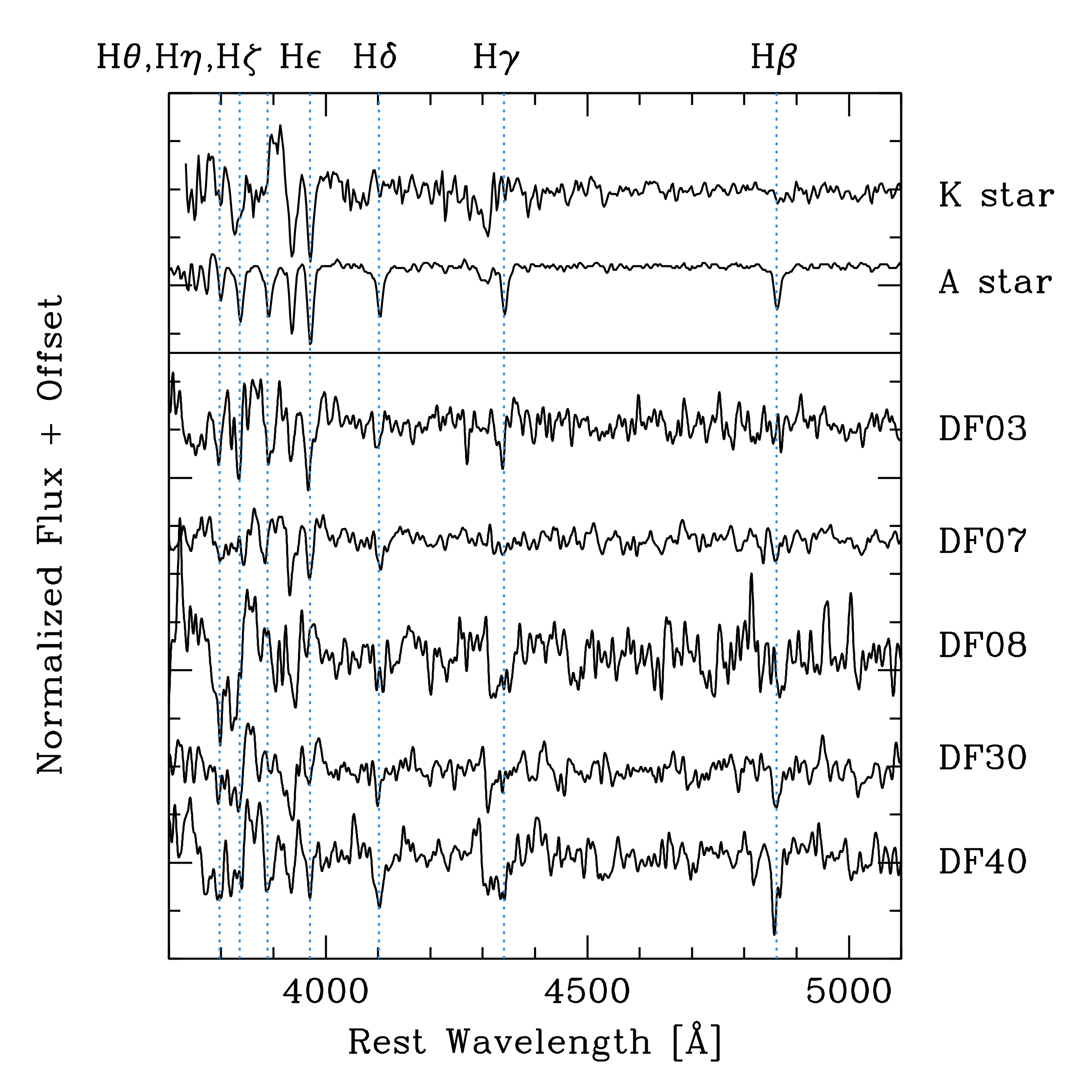}
\caption{Rest-frame, continuum-subtracted spectra of the observed UDGs for which we can measure a reliable redshift. Figure 4 highlights the oxygen emission lines in DF08. We include the A and K-type continuum-subtracted stellar spectra for comparison. We label the Balmer lines to guide the eye.
}
\label{fig:udg_spectra}
\end{figure}

\section{Results}
\label{results}

We now describe our measurement of the individual recessional velocities of the UDGs and examine the nature of the spectra, both in terms of the individual spectra and a combined stack. We confine our discussion to the blue channel spectra (Figure \ref{fig:udg_spectra}). 
The most prominent absorption features in the UDG spectra are the Balmer absorption lines, the Ca II H \& K lines just blueward of restframe 4000 \AA\, and the G-band at about 4300 \AA. These features are also prominent in the published spectrum of DF44 \citep{vanDokkum2015b}.

\subsection{Redshifts and Coma Cluster Membership}

We measure redshifts and uncertainties by cross-correlating the UDG spectra against spectral templates. 
The ubiquity of Balmer lines suggests that an A-type stellar spectrum would be an ideal template, although the Ca lines and G-band suggest that later type stars might also be suitable. As such, we use both, but generally find more significant results using the A-type template, with velocities differing by no more than 60 km s$^{-1}$ between templates in most cases, with one exception at 120 km s$^{-1}$.

We use the IRAF\footnote{IRAF is distributed by the National Optical Astronomy Observatory, which is operated by the Association of Universities for Research in Astronomy (AURA) under a cooperative agreement with the National Science Foundation.} task XCSAO for the cross-correlation analysis. We vary parameters, including the high and low frequency cutoffs,
while searching velocities between $-5000 \, \mathrm{km \, s^{-1}}$ and $30,000 \, \mathrm{km \, s^{-1}}$) for the correlation function maximum. We only accept velocities that are robust to changes in the input parameters and then visually confirm that multiple absorption lines correspond to the principal absorption features that we mention above. We extract high confidence redshifts from the spectra of 5 out of the 6 observed UDGs. For the sixth, we do not even extract a low confidence redshift estimate. 

Because of heliocentric velocity corrections, differences in slit illumination, and wavelength calibration errors, it is possible for template stars to have effective reference velocities that differ by several tens of km s$^{-1}$ from the published values. Such uncertainties are irrelevant when determining whether galaxies lie within the Coma cluster, but the surprising availability of emission lines in one of our spectra allows us to correct the velocity zero point and also provides a test of the absorption line cross-correlation procedure. We will discuss this interesting object further below.

We compare emission and absorption redshifts for DF08 to estimate the external uncertainties of the stellar cross correlation analysis. We find an offset of 220 km s$^{-1}$ between the recessional velocities measured from cross-correlation with spectral templates and from the shift in [OII] and [OIII] lines with restframe wavelengths in air of 3726.05, 3728.80, 4958.92, 5006.85 \AA\ (about 2$\sigma$ discrepant given the internal error estimate). Differences among the cross-correlation results using different templates for all of our UDGs are $<$ 120 km s$^{-1}$. We conclude that even if our uncertainties are truly as large as 200 km s$^{-1}$, our conclusions regarding membership in Coma for these UDG candidates is unchanged.

We convert the spectra of the five UDGs from observed to rest frame, subtract the continua, and display them in Figure \ref{fig:udg_spectra}. The best-fit UDG recessional velocities and uncertainties are listed in Table \ref{table:udg_target}.

\begin{figure}[t]
\includegraphics[width=0.5\textwidth]{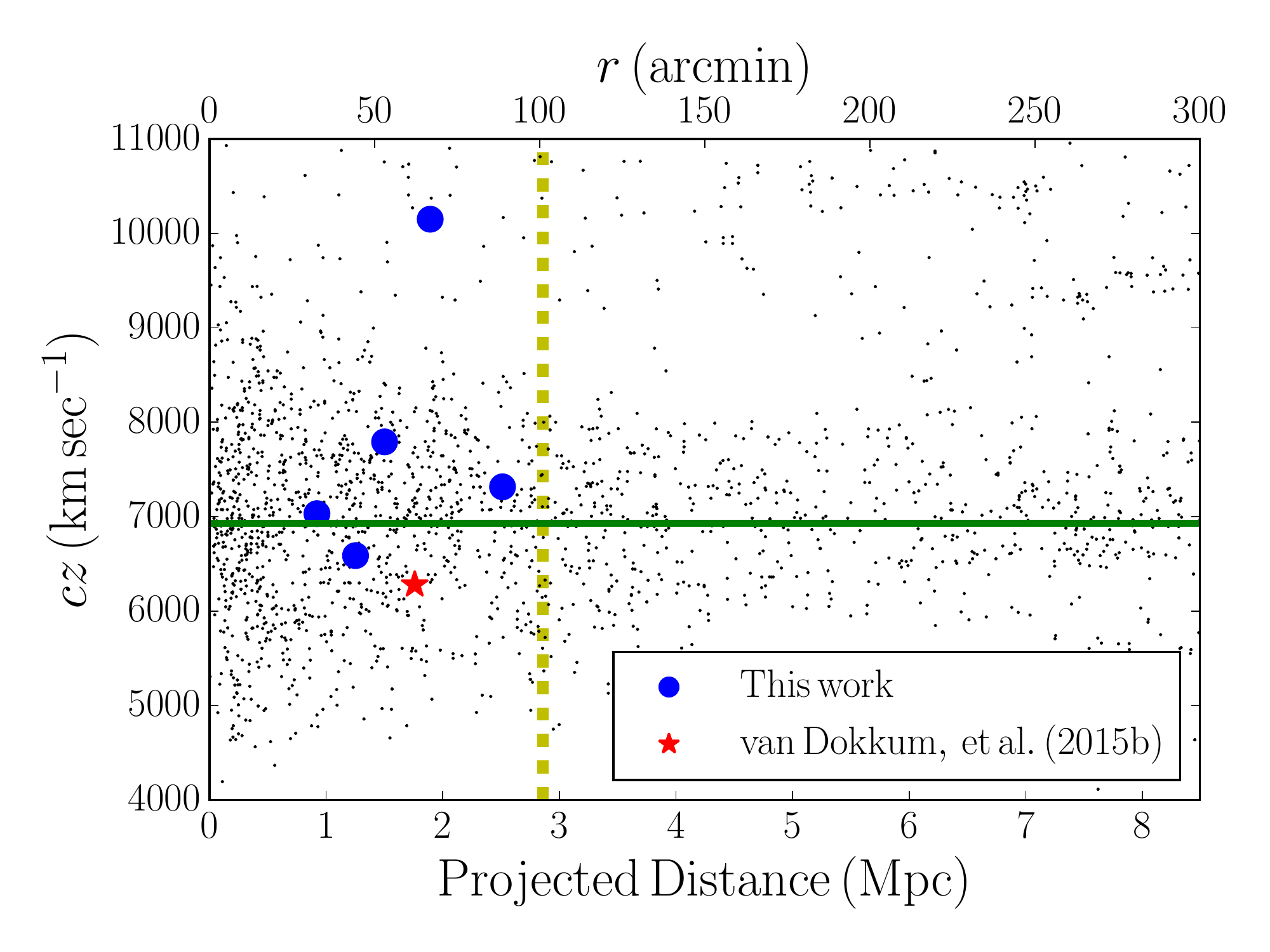}
\caption{Phase space diagram of the Coma cluster. The small black dots represent individual galaxies projected near the cluster, which mostly form the characteristic caustic pattern centered on Coma's mean recessional velocity (horizontal green line). The dashed yellow line indicates Coma's virial radius \citep{Kubo2007}. The large red star represents the location of DF44 \citep{vanDokkum2015a}, with uncertainties smaller than the symbol. The large blue circles represent the locations of our UDGs. We conclude that four UDGs from our sample (DF07, DF08, DF30, DF40) and DF44 are cluster members and that DF03 is well behind the Coma cluster.}
\label{fig:phase_space}
\end{figure}

To determine Coma cluster membership, we place the UDGs on the cluster phase-space diagram, which we construct using measurements of galaxies retrieved from the NASA Extragalactic Database (Figure \ref{fig:phase_space}). There are enough known Coma members that they clearly delineate the classic wedge distribution defined by the caustic curves. Four of our UDGs (DF07, DF08, DF30, and DF40), plus the previously measured DF44, fall within the caustics and we conclude these are bona fide members. Another of our UDGs (DF03) lies well outside the caustic curves and we conclude that this system is 45 Mpc behind the cluster. From the statistics that 5 out of the 6 UDGs with measured redshifts lie within Coma and restricting ourselves to systems that satisfy our selection criteria, we conclude that of the 18 \cite{vanDokkum2015a} systems that satisfy our size and magnitude criteria, $\sim 15$ are likely to be Coma cluster members. We conclude that there is a significant population of physically large cluster UDGs.
We do not interpret the phase-space diagram further because the radial selection effects in the original catalog are likely to be significant. 


\begin{figure}[t]
\epsscale{1.0}
\plotone{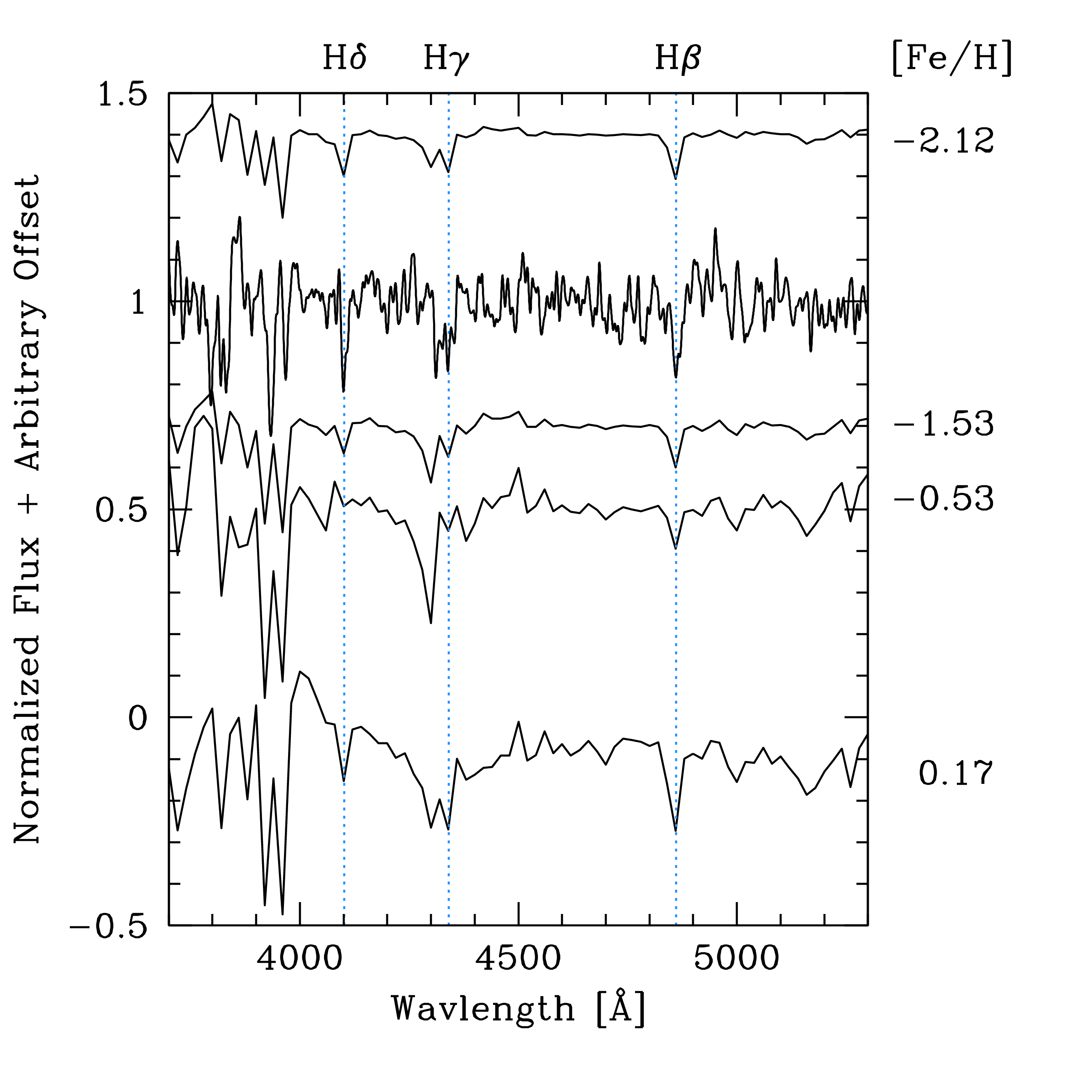}
\caption{A comparison of the UDG composite spectrum (second from the top) and modeled stellar population spectra of differing metallicity and age (continuum subtracted in all cases). The composite includes all five of our UDGs. We have placed it, on the basis of visual inspection, in order among the metallicity sequence on the basis of the H$\delta$ absorption depth and the ratio of the G-band to H$\gamma$. The model spectra with [Fe/H] $< -0.5$ are all old (13 Gyr) populations, while the metal rich spectrum has been optimized to provide the best fit for a 1 Gyr population.}
\label{fig:stacked_metallicity}
\end{figure}

\subsection{Stellar Populations and Metallicity}

The presence of Balmer lines in galaxy spectra is often interpreted as evidence for intermediate age (1 Gyr) populations, but can also be the
result of low metallicity. We suspect the latter is the case in these galaxies given their red colors as a population \citep{Koda2015}. 
To explore this issue further, we  
combine the spectra to produce a higher signal-to-noise spectrum (Figure \ref{fig:stacked_metallicity}), but caution that this approach caries a risk that we are combining
systems with different properties. 

We use the P\'EGASE stellar population modeling software \citep{pegase} to create comparison spectra for single age (13 Gyr) populations with three different metallicities (Figure \ref{fig:stacked_metallicity}). We visually place the stacked spectrum in the abundance sequence. In particular, we focus on the strength of the H$\delta$ line and the ratio of the H$\gamma$ line to the G-band, which lies just to the blue of H$\gamma$. This analysis is clearly a preliminary determination of the typical metallicity of such UDGs, and a composite one at that, but the indications are that these are quite metal poor ([Fe/H] $\lesssim -1.5$). 

For comparison, we also identify the optimal match for a young, 1 Gyr, population in Figure \ref{fig:stacked_metallicity}. We prefer the models with old, metal poor populations for several reasons. First, the spectrum of the young population matches H$\beta$ and H$\gamma + $G-band, but results in an H \& K break that is too strong, as seen by the peak at around 4000\AA, and a Mg and Fe complex at about 5200\AA\ that is not in the observed spectra. Second, the low metallicity estimate for these galaxies agrees with a simple expectation from the metallicity-luminosity relation \citep[][(ZKH)]{Zaritsky1994}. 
For early type galaxies, $g-B$ is roughly $-0.8$ \citep{Fukugita}, which places our typical UDG with $M_g \sim -15$ at $M_B \sim -14.2$ and at about
an [Fe/H] of $-1.5$ in the relationships plotted in Figure 13 of ZKH. 
Finally, independent measurements of the metallicity in similar galaxies has also concluded that these systems are of low metalliity \citep{makarov2015}.
We conclude, based on the low metallicites, that these UDGs are not the tidal remnants of much larger galaxies.

\begin{figure}
\epsscale{1.0}
\plotone{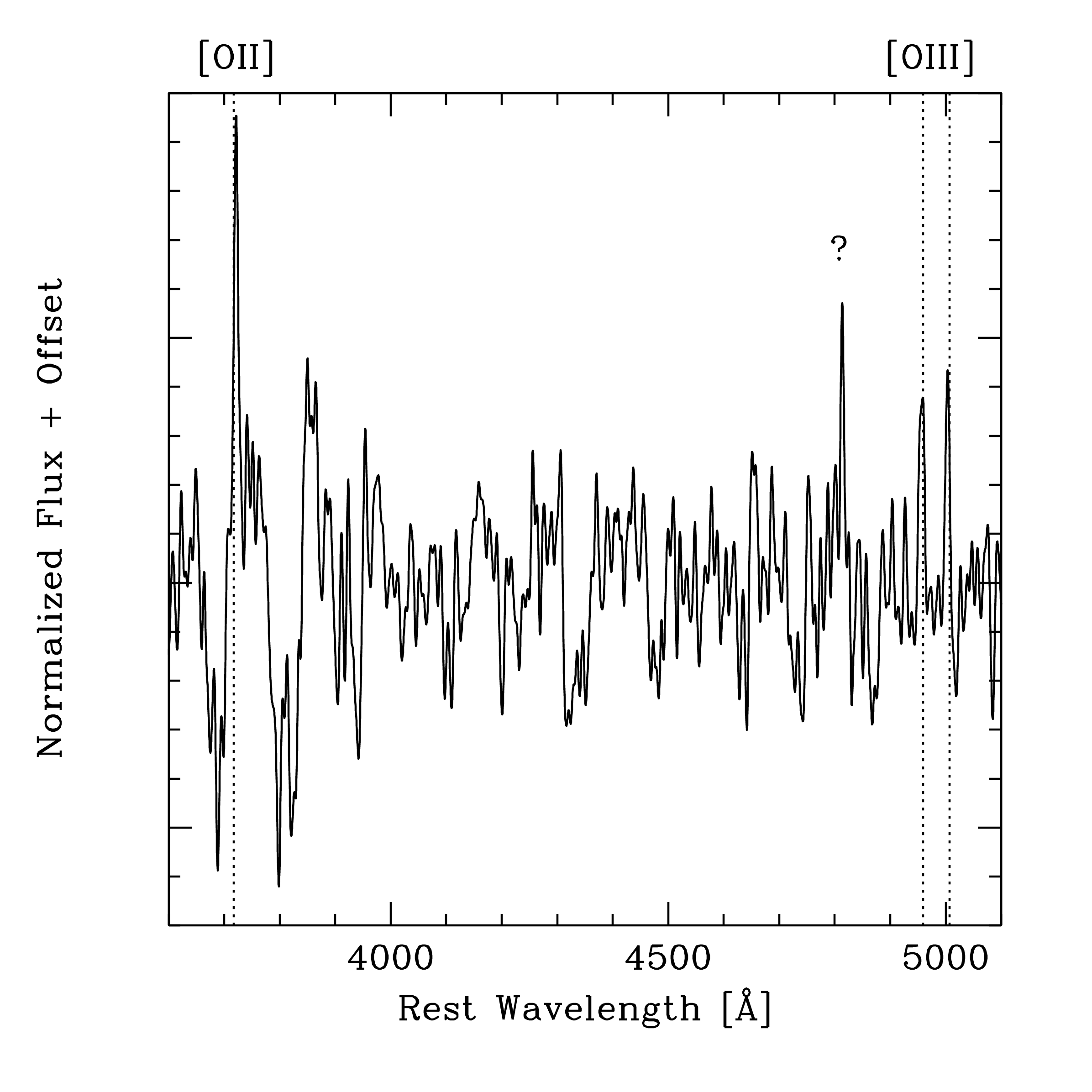}
\caption{Smoothed spectrum of DF08, which shows [OII] and [OIII] emission lines at the recessional velocity derived from the absorption lines. This is the first cluster UDG spectrum with emission lines, demonstrating that some UDGs, even in the cluster environment, retain some of their gas and host ionizing sources. The apparent emission line labeled `?' is interpreted as noise. It is of comparable magnitude to the apparent, but non physical, absorption features near it. This level of noise is also responsible for the deviation of the 4959/5007 line flux of 1:3.}
\label{fig:OII_OIII}
\end{figure}

\subsection{Emission Lines}
While absorption features are common in previously observed UDG spectra \citep{vanDokkum2016, Martinez-Delgado2016}, emission lines have been found only in one gas rich UDG \citep{trujillo2017}.
Surprisingly, we have a weak detection of oxygen in one of the Coma UDGs in our sample.
In Figure \ref{fig:OII_OIII} we show a spectrum of DF08, which 
features weak [OII] and [OIII] emission lines at a redshift consistent with the absorption line redshift. 
The existence of these lines indicates that DF08 is not entirely gas-depleted nor devoid of ionizing sources.
We are unable to localize a specific source in the 2D spectrum, so we conclude that the emission is somewhat physically extended. Emission lines may become a more common signature of UDGs as field samples are explored \citep{DiCintio2017, trujillo2017} and establishing the differences between field and cluster UDG samples is a clear next step.


\section{Conclusions}
We present spectroscopy of five UDGs seen in projection on the Coma Cluster with the MODS spectrograph on the LBT and reach the following conclusions. 

\begin{itemize}
\item On the basis of their recessional velocities, we confirm 4 of our UDGs to be Coma cluster members, thereby quintupling the population of spectroscopically confirmed Coma UDGs. 

\item On the basis of its recessional velocity, we place another of our UDGs about 45 Mpc behind the Coma cluster.
Along with DGSAT I, DF03 is one of the few field UDGs with a spectroscopic redshift. 

\item Coupled with the spectroscopic confirmation of DF44 \citep{vanDokkum2015b}, the result that 5 of 6 spectroscopically observed, physically large (projected half light radius $>$ 2.9 kpc) UDGs from the \cite{vanDokkum2015a} catalog are bona fide cluster members suggests that $\sim 15$ of the 18 similarly large Dragonfly UDGs are Coma members.

\item On the basis of a comparison between P\'{E}GASE stellar population synthesis models and our composite UDG spectra, we conclude that, on average, these systems are metal-poor ([Fe/H] $\lesssim -1.5$). These systems are consistent with the metallicity-luminosity relation and  the result excludes the possibility that these systems are the relics of much more luminous galaxies. 

\item We present the first cluster UDG (DF08) spectrum with emission lines. This finding demonstrates that not all cluster UDGs lack gas and sources of ionizing radiation.

\end{itemize}



\acknowledgments
The authors made use of the online cosmology calculator as described in \cite{Wright2006}.

The authors thank Barry Rothberg, our LBT support astronomer, and Olga Kuhn who helped us to navigate through the MODS data reduction pipeline and to acquire the necessary blue and red spectral flats for our custom slit.

The LBT is an international collaboration among institutions in the United States, Italy and Germany. LBT Corporation partners are: The University of Arizona on behalf of the Arizona university system; Istituto Nazionale di Astrofisica, Italy; LBT Beteiligungsgesellschaft, Germany, representing the Max-Planck Society, the Astrophysical Institute Potsdam, and Heidelberg University; The Ohio State University, and The Research Corporation, on behalf of The University of Notre Dame, University of Minnesota and University of Virginia.
This paper uses data taken with the MODS spectrographs built with funding from NSF grant AST-9987045 and the NSF Telescope System Instrumentation Program (TSIP), with additional funds from the Ohio Board of Regents and the Ohio State University Office of Research.
The NASA/IPAC Extragalactic Database (NED) is operated by the Jet Propulsion Laboratory, California Institute of Technology, under contract with the National Aeronautics and Space Administration.

{\it Facilities:} \facility{LBT (MODS)}.





\clearpage

\end{document}